\documentclass[12pt,preprint]{aastex}

\slugcomment{To appear in AJ}

\shorttitle{Spectral Variability of PKS 1718-649}
\shortauthors{Tingay et al.}

\begin{document}


\title{The Spectral Variability of the GHz-Peaked Spectrum Radio Source PKS 1718$-$649 and a Comparison of Absorption Models}


\author{S.J.~Tingay\altaffilmark{1},
J.-P.~Macquart\altaffilmark{1},
J.D.~Collier\altaffilmark{2,3},
G.~Rees\altaffilmark{4,3},
J.R.~Callingham\altaffilmark{5,6},
J.~Stevens\altaffilmark{3},
E.~Carretti\altaffilmark{3},
R.B.~Wayth\altaffilmark{1},
G.F.~Wong\altaffilmark{2,3},
C.M.~Trott\altaffilmark{1,6},
B.~McKinley\altaffilmark{7},
G.~Bernardi\altaffilmark{8},
J.~D.~Bowman\altaffilmark{9}, 
F.~Briggs\altaffilmark{7},
R.~J.~Cappallo\altaffilmark{10}, 
B.~E.~Corey\altaffilmark{10}, 
A.~A.~Deshpande\altaffilmark{11}, 
D.~Emrich\altaffilmark{1},
B.~M.~Gaensler\altaffilmark{5,6},
R.~Goeke\altaffilmark{12},
L.~J.~Greenhill\altaffilmark{13},
B.~J.~Hazelton\altaffilmark{14}, 
M.~Johnston-Hollitt\altaffilmark{15},
D.~L.~Kaplan\altaffilmark{16}, 
J.~C.~Kasper\altaffilmark{17},
E.~Kratzenberg\altaffilmark{10}, 
C.~J.~Lonsdale\altaffilmark{10}, 
M.~J.~Lynch\altaffilmark{1}, 
S.~R.~McWhirter\altaffilmark{10},
D.~A.~Mitchell\altaffilmark{3,6}, 
M.~F.~Morales\altaffilmark{14}, 
E.~Morgan\altaffilmark{12}, 
D.~Oberoi\altaffilmark{18}, 
S.~M.~Ord\altaffilmark{1},
T.~Prabu\altaffilmark{11}, 
A.~E.~E.~Rogers\altaffilmark{10}, 
A.~Roshi\altaffilmark{19}, 
N.~Udaya~Shankar\altaffilmark{11}, 
K.~S.~Srivani\altaffilmark{11}, 
R.~Subrahmanyan\altaffilmark{11,6},
M.~Waterson\altaffilmark{1}
R.~L.~Webster\altaffilmark{20,6}, 
A.~R.~Whitney\altaffilmark{10}, 
A.~Williams\altaffilmark{1} 
C.~L.~Williams\altaffilmark{12}}

\altaffiltext{1}{International Centre for Radio Astronomy Research, Curtin University, Bentley, WA 6102, Australia}
\altaffiltext{2}{University of Western Sydney, Locked Bag 1797, Penrith, NSW 2751}
\altaffiltext{3}{CSIRO Astronomy and Space Science (CASS), PO Box 76, Epping, NSW 1710, Australia}
\altaffiltext{4}{Macquarie University, Balaclava Rd, North Ryde, NSW 2109, Australia}
\altaffiltext{5}{Sydney Institute for Astronomy, School of Physics (SiFA), The University of Sydney, NSW 2006, Australia}
\altaffiltext{6}{ARC Centre of Excellence for All-sky Astrophysics (CAASTRO)}
\altaffiltext{7}{Research School of Astronomy and Astrophysics, Australian National University, Canberra, ACT 2611, Australia}
\altaffiltext{8}{Square Kilometre Array South Africa (SKA SA), Cape Town 7405, South Africa}
\altaffiltext{9}{School of Earth and Space Exploration, Arizona State University, Tempe, AZ 85287, USA}
\altaffiltext{10}{MIT Haystack Observatory, Westford, MA 01886, USA}
\altaffiltext{11}{Raman Research Institute, Bangalore 560080, India}
\altaffiltext{12}{Kavli Institute for Astrophysics and Space Research, Massachusetts Institute of Technology, Cambridge, MA 02139, USA}
\altaffiltext{13}{Harvard-Smithsonian Center for Astrophysics, Cambridge, MA 02138, USA}
\altaffiltext{14}{Department of Physics, University of Washington, Seattle, WA 98195, USA}
\altaffiltext{15}{School of Chemical \& Physical Sciences, Victoria University of Wellington, Wellington 6140, New Zealand}
\altaffiltext{16}{Department of Physics, University of Wisconsin--Milwaukee, Milwaukee, WI 53201, USA}
\altaffiltext{17}{Department of Atmospheric, Oceanic and Space Sciences, University of Michigan, Ann Arbor, MI 48109, USA}
\altaffiltext{18}{National Centre for Radio Astrophysics, Tata Institute for Fundamental Research, Pune 411007, India}
\altaffiltext{19}{National Radio Astronomy Observatory, Charlottesville and Greenbank, USA}
\altaffiltext{20}{School of Physics, The University of Melbourne, Parkville, VIC 3010, Australia}

\email{s.tingay@curtin.edu.au}

\begin{abstract}
Using the new wideband capabilities of the Australia Telescope Compact Array (ATCA), we obtain spectra for PKS 1718$-$649, a well-known gigahertz-peaked spectrum radio source.  The observations, between approximately 1 and 10 GHz over three epochs spanning approximately 21 months, reveal variability both above the spectral peak at $\sim$3 GHz and below the peak.  The combination of the low and high frequency variability cannot be easily explained using a single absorption mechanism, such as free-free absorption or synchrotron self-absorption.  We find that the PKS 1718$-$649 spectrum and its variability are best explained by variations in the free-free optical depth on our line-of-sight to the radio source at low frequencies (below the spectral peak) and the adiabatic expansion of the radio source itself at high frequencies (above the spectral peak).  The optical depth variations are found to be plausible when X-ray continuum absorption variability seen in samples of Active Galactic Nuclei is considered.  We find that the cause of the peaked spectrum in PKS 1718$-$649 is most likely due to free-free absorption.  In agreement with previous studies, we find that the spectrum at each epoch of observation is best fit by a free-free absorption model characterised by a power-law distribution of free-free absorbing clouds.  This agreement is extended to frequencies below the 1 GHz lower limit of the ATCA by considering new observations with Parkes at 725 MHz and 199 MHz observations with the newly operational Murchison Widefield Array. These lower frequency observations argue against families of absorption models (both free-free and synchrotron self-absorption) that are based on simple homogenous structures.
\end{abstract}

\keywords{galaxies: active -- galaxies: individual (PKS 1718$-$649, NGC 6328) -- radio continuum: galaxies -- radiation mechanisms: general}

\section{Introduction}
PKS 1718$-$649 (NGC 6328: $z=0.014428\pm0.000023$ from \citep{mey04}) is a well-studied gigahertz-peaked spectrum (GPS) radio galaxy, displaying all the classic attributes of this class of object, as presented in previous studies \citep{tin97,tin03}: a compact double radio morphology on parsec scales; low radio variability; low radio polarisation; a persistently peaked radio spectrum; and a kinematically complicated, high gas density host galaxy environment.  Recently the GPS classification has been supplanted by the Compact Symmetric Object (CSO) classification, where CSOs are suggested to be young radio galaxies and possess the characteristics that PKS 1718$-$649 displays \citep{wil94}.  The CSO classification guards against the selection of radio AGN that occasionally possess peaked radio spectra \citep{tre11}.

The origin of the peaked radio spectrum in this class of object is still a matter for debate, with a number of absorption models previously proposed to explain the inverted low frequency radio spectrum.  \citet{tin03} reviewed the various models and tested many of them against radio data obtained for PKS 1718$-$649 from the Australia Telescope Compact Array (ATCA), consisting of 30 - 40 flux density measurements over the frequency range 1 - 9 GHz over four epochs of observation spanning 14 months.  The goal of these observations was to characterise the radio spectrum of PKS 1718$-$649 and attempt to use the radio spectrum and its spectral variability (spectral variability in this paper is defined as the variability of the spectral shape as a function of time) to constrain the different absorption models.  \cite{tin03} concluded that the leading contenders, free-free absorption and synchrotron self-absorption, produced similar quality parameterisations of the measured spectra, and that the spectral variability observed was marginal and difficult to interpret given the measurement errors.  Although an excellent fit to the data was obtained with the free-free absorption model proposed by \citet{bic97}, the spectral coverage of the measurements at low frequencies did not allow a compelling test of the model.  Taking all of the evidence into account, including VLBI measurements of the sizes of the compact radio emitting regions, the conclusion of \cite{tin03} was that synchrotron self-absorption was the most likely contributor to the peaked radio spectrum in PKS 1718$-$649.

While the \cite{tin03} study represented a very detailed investigation of absorption models for PKS 1718$-$649, the measurements were sub-optimal in some respects.  The spectral coverage of the ATCA at the time of the observations was good at frequencies within the 6 cm and 3 cm bands but sparse at frequencies in the 21 cm and 13 cm bands.  At 21 cm and 13 cm, due to the presence of Radio Frequency Interference (RFI) and the limitation of the 128 MHz instantaneous bandwidth of the ATCA, it was difficult to obtain good frequency coverage below the spectral peak for PKS 1718$-$649 (at approximately 3 GHz, between the 13 cm and 6 cm bands).  The paucity of the low frequency coverage, where absorption processes have the largest effect, somewhat hindered the investigation.

Due to the installation of the new Compact Array Broadband Backend (CABB) \citep{wil11}, a far better instantaneous frequency coverage is now available in all bands of the ATCA, meaning that monitoring observations focused on spectral coverage are far easier and more robust than a decade ago.  An example of the quality and utility of results possible with the CABB system can be seen in \citet{mac13}, where broadband measurements aided the study of microarcsecond-scale structure in AGN via intersteller scintillation.  The new CABB system has motivated us to revisit PKS 1718$-$649, in order to better characterise its spectrum and spectral variability.  The new CABB data allow a better {\bf measurement} of the spectrum and also reveal other possible issues with the previous study by \citet{tin03}.

In addition, the new low frequency radio telescope, the Murchison Widefield Array (MWA) \citep{tin13}, is starting to produce very large surveys of the southern sky in continuum radio emission between the frequencies of 80 and 300 MHz, substantially lower than the lowest frequency available at the ATCA.  We use the MWA, as well as the Parkes radio telescope, to help characterise the PKS 1718$-$649 spectrum well below the peak frequency, which assists in distinguishing between absorption models.

\section{Observations and data analysis}

\subsection{Australia Telescope Compact Array observations}

In 2012-2013, a series of four snapshot observations of PKS 1718$-$649 were made using the ATCA during unallocated ``green time". The CABB system gives an instantaneous 2 GHz bandwidth for both linear polarisations. The observations are summarised in Table~\ref{observationParams}. Frequency switching was used between the 16$\,$cm (2.1 GHz) and 4$\,$cm (5.5 and 9.0 GHz) bands, which required adjusting the subreflector focus for some antennas. The frequency ranges for each of these bands were 1075 $-$ 3123 MHz, 4477 $-$ 6525 MHz, and 7977 $-$ 10025 MHz, respectively.  Observations for all bands used 2048$\times$1 MHz channels and a 10 second correlator integration time. For each observation, PKS 1934$-$638 was used for bandpass, flux density, and phase calibration, since it is only $\sim$15 degrees away from PKS 1718-649. Each observation repeated through a cycle of 10 minute scans of PKS 1718$-$649 bracketed by 5 minute scans of PKS 1934$-$638, before switching frequencies and repeating the process.

\begin{table}
\begin{center}
\caption{Summary of the ATCA observations.}
\begin{tabular}{ccccc}
\tableline\tableline
Project ID & Date & Frequency (GHz) & Duration (hours) & Array Configuration\\
\tableline
C1034 & 2012 Feb 29 & 2.1, 5.5, 9.0 & 0.9 & 750D\\
C1034 & 2012 Sep 27 & 2.1, 5.5, 9.0 & 2.6 & H214\\
C2768 & 2013 Feb 17 & 2.1, 5.5, 9.0 & 2.5 & 6A\\
C2768 & 2013 Dec 16 & 2.1, 5.5, 9.0 & 1.1 & 750B\\

\tableline
\label{observationParams}
\end{tabular}
\end{center}
\end{table}

The \textsc{miriad} software package \citep{sau95} was used to analyse the data. Each set of observations was analysed using the same process, as follows.

The data were loaded using {\tt atlod}.  The 40 channels at either end of the band were flagged using {\tt uvflag}.  The data were split into single observing band datasets using {\tt uvsplit}.  We iteratively flagged and calibrated PKS 1934$-$649, checking flagging with {\tt uvspec} and calibration with {\tt uvplt}.  Data were flagged for RFI using {\tt pgflag} and {\tt blflag}.  Calibration consisted of estimating the bandpass solution using {\tt mfcal} and then estimating the gain (amplitude and phase) and leakage solutions over four frequency bins and a solution interval of 10 seconds using {\tt gpcal}.  This flagging and calibration process was repeated until good calibration solutions were found.  The calibration solutions were transferred to PKS 1718$-$649 using {\tt gpcopy}.  The PKS 1718$-$649 data were then flagged and calibrated, also solving for the source polarisation in {\tt gpcal}.  The flux density scale was set to that defined by PKS 1934$-$638 by using {\tt gpboot}.

During the analysis, the 2012 September 27 data were found to have had a number of instrumental problems during the observations, such that no stable calibration solutions could be found. These data were therefore discarded and further observations were obtained on 2013 February 17.

The excellent frequency coverage provided by CABB allows a closer examination of the spectrum of PKS 1718$-$649 than was possible by \cite{tin03}, especially at low frequencies.  This is illustrated in Figure 1, which shows the calibrated Stokes I measurements from 2013 December 16 on a single baseline (antennas 3 - 4; 415 m length baseline) as a function of frequency.  Clearly apparent in Figure 1 is significant beating in the visibility amplitude, indicating the presence of a confusing source {\bf $\sim$40\arcmin} from PKS 1718$-$649 on the sky (identified as 1722$-$64: RA=17:26:57.813; DEC=$-$64:27:52.79; J2000.  Approximately 4 Jy at 16 cm and 1 Jy at 6 cm\footnote{http://www.narrabri.atnf.csiro.au/calibrators (1722-64 does not appear in the Parkes catalog)}).  Such confusing structure was not apparent in the data of \cite{tin03}, due to limited spectral coverage.  The effect of the confusing source changes with frequency (both due to the response of the interferometer and the response of the individual antenna elements), baseline length, and orientation.

The confusing object needs to be identified, modelled, and its effects removed from the data in the ($u,v$) plane before the spectrum of PKS 1718$-$649 can be isolated and measured.  To model the confusing object, we used the {\tt uvsfit} task in \textsc{miriad}.  {\tt uvsfit} fits model components to visibility data, taking into account the frequency dependence of the model component flux densities.  {\tt uvsfit} was run at each epoch by simultaneously fitting all available data at all frequencies, using two point source model components (PKS 1718$-$649 and 1722$-$64 are both unresolved at the frequencies used here with the ATCA), and allowing {\tt uvsfit} full freedom to utilise the frequency dependence of the flux density for both components.  {\tt uvsfit} produces visibility datasets representing the fitted model, visibility datasets containing the residuals on the fits, and reports the parameters of the models with errors on those parameters.  Figure 1 shows the model fit to the data from the December 2013 observation.

The model parameters generated for each epoch of observation are given in Table 2.  The reader is referred to the documentation for {\tt uvsfit}\footnote{http://www.atnf.csiro.au/computing/software/miriad/doc/uvsfit.html} for an explanation of the model parameters (and the caption to Table 2).  For all of the model fits performed, a reference frequency of 3.0 GHz was chosen in {\tt uvsfit}.  The RMS residuals resulting from the fits in February 2012, February 2013 and December 2013 were 98 mJy, 69 mJy, and 64 mJy, respectively.  The component representing PKS 1718$-$649 in the model fits was held constant at the interferometer phase centre.  The 1722$-$64 position was not held constant, to allow for any phase errors due to the differential ionosphere and/or atmosphere to be accounted for in the fit.  The expected angular separation between PKS 1718$-$649 and 1722$-$64 is 0.6476$^{\circ}$ and the measured angular distance from Table 2 is 0.6470$^{\circ}$.

\begin{figure}[ht]
\begin{center}
\includegraphics[width=12cm]{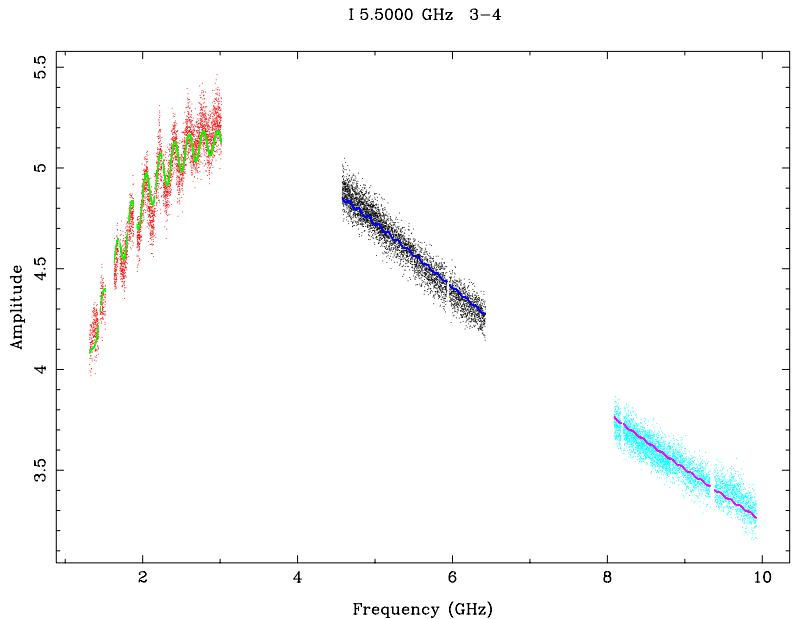}
\caption{Measured spectrum for PKS 1718$-$649 from the December 2013 data on a single 415 m baseline, shown with a model fit derived via MIRIAD task {\tt uvsfit}, as described in the text.  Red is the 16 cm (2.1 GHz) band data, with green as model.  Black is the 4 cm (5.5 GHz) band data, with dark blue as model.  Light blue is the 4 cm (9.0 GHz) band data, with red as model.  The data, particularly in the 16 cm band show the beating of the visibility amplitude, indicating a confusing source.}
\end{center}
\end{figure}

\begin{table}[ht]
\begin{center}
\begin{tabular}{ l c  c c c c c} \hline
Date  & Source & Flux density      & Offset position                   & $\alpha_{0}$      & $\alpha_{1}$         & $\alpha_{2}$  \\
      &           &(Jy)               & RA/DEC (\arcsec)                  &                   &                     &               \\ \hline
Feb 2012& A&5.194$\pm$0.002  &0/0                                &0.0881$\pm$0.0001  &$-$0.3812$\pm$0.0001 &0.0210$\pm$0.0001 \\
Feb 2012&B  &0.036$\pm$0.003   &1272.12$\pm$0.11/  &$-$2.313$\pm$0.002 &$-$1.630$\pm$0.002   &2.311$\pm$0.003 \\
&&&1950.99$\pm$0.16&&& \\
Feb 2013& A&5.222$\pm$0.001  &0/0                                &0.0143$\pm$0.0001  &$-$0.3580$\pm$0.0001 &0.0211$\pm$0.0001 \\
Feb 2013&B  &0.038$\pm$0.001   &1276.14$\pm$0.05/  &$-$2.124$\pm$0.007 &$-$1.339$\pm$0.006   &2.029$\pm$0.011 \\
&&&1954.71$\pm$0.08&&& \\
Dec 2013& A&5.130$\pm$0.002  &0/0                                &0.0068$\pm$0.0001  &$-$0.3318$\pm$0.0002 &0.0093$\pm$0.0002 \\
Dec 2013&B  &0.054$\pm$0.001   &1256.16$\pm$0.20/    &$-$2.87$\pm$0.03   &$-$2.10$\pm$0.02     &2.40$\pm$0.04 \\ 
&&&1887.24$\pm$0.20&&& \\ \hline
\end{tabular}
\caption{The models produced from {\tt uvsfit} for each of the three datasets.  Source A is PKS 1718$-$649.  Source B is 1722$-$64.  $\alpha_{0}$, $\alpha_{1}$, and $\alpha_{2}$ are defined in $\alpha(\nu)=\alpha_{0}+l_{fr}(\alpha_{1}+l_{fr}\alpha_{2})$, where $\alpha(\nu)$ is the spectral index as a function of frequency.  $l_{fr}=ln(\nu/\nu_{0})$, where $\nu_{0}$ is the reference frequency.  The flux density in the table is at the reference frequency (3.0 GHz for these models).}
\end{center}
\end{table}

Errors on the spectral model for PKS 1718$-$649, for each spectral channel, for each observation, are determined as follows.  The RMS residuals between the measured visibility amplitudes and the model visibility amplitudes, for each spectral channel, over each observation duration, are taken as the one sigma errors on the individual spectral channels.  Some variation of these errors over the fitted frequency change is apparent, with larger errors assigned to parts of the frequency range that are not as well fit by the model as other frequency ranges.  Also, a relatively conservative RFI flagging has been applied, causing a small number of spectral channels to have larger than typical errors due to residual RFI.  Figure 2 shows the model spectra for PKS 1718$-$649 for the three observations, from Table 2, along with the errors on the flux densities per spectral channel.

\begin{figure}[ht]
\begin{center}
\includegraphics[width=12cm]{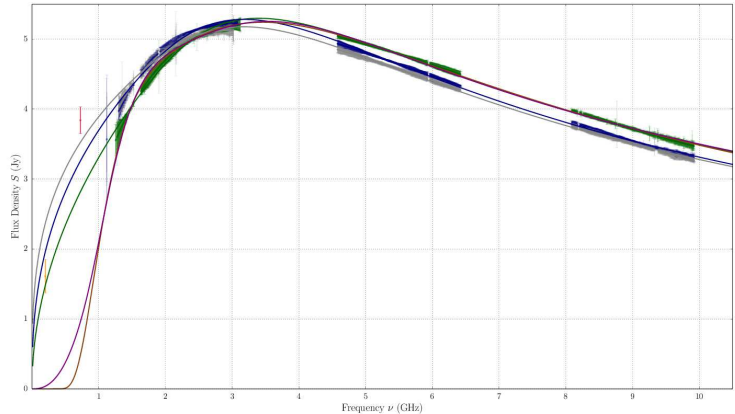}
\caption{Measured radio spectra for PKS 1718$-$649 as described in the text, for the three epochs of observation.  Also shown are a variety of model fits to the spectra, as described in the text.  The ATCA data for 2012 February are shown in green.  The ATCA data for 2013 February are shown in blue.  The ATCA data for 2013 December are shown in grey.  The fits of the \cite{bic97} model to these data are shown as solid lines in the same colours.  Also shown are example homogenous free-free and synchrotron self-absorption model fits to the 2012 February data: purple solid line for a homogenous synchrotron self-absorption model with two components; brown solid line for a homogenous free-free absorption model with two components.  Finally, data from Parkes at 725.75 MHz and the MWA at 199 MHz are shown, but not used in the fitting, as described in the text.}
\end{center}
\end{figure}

The model parameters for 1722$-$64 are difficult to interpret, as they represent the true spectrum of the radio source as well as the frequency dependent response of the antennas beyond the half power points.

We note that we do not explore the polarisation properties of PKS 1718$-$649 in this paper, as our sensitivity to polarised emission is lower than the deep polarisation observations of \cite{tin03}.  Spectro-polarimetry of PKS 1718$-$649 could provide several interesting tests of the absorption mechanisms under consideration here \citep{ino02} but would require deep CABB observations that are properly configured for accurate polarisation calibration.  This will be the subject of a future investigation.

In \cite{tin03}, the error analysis took into account the errors on the absolute flux density scale in the Southern Hemisphere, strictly rendering the errors on the flux density measurements comparable to the flux density variations observed.  In order to utilise the current data for variability analyses, here we assume that the error on the flux density scale in the Southern Hemisphere remains constant over the course of the observations.  Explicit in this assumption is that the spectrum of the ATCA primary flux density calibrator, PKS 1934$-$638, does not change over the course of the observations.  This assumption is supported by the results of multi-frequency, multi-epoch monitoring of a sample of stable, compact radio sources by \cite{tin03b}, showing that epoch-to-epoch variations of PKS 1934$-$638 over sixteen epochs spanning three and a half years were limited to less than 2\%.  The variations in the PKS 1718$-$649 flux densities shown in Figure 2 significantly exceed 2\%.  We therefore ascribe these variations to {\bf PKS 1718--649} itself, rather than any variability in {\bf PKS 1934--638}.

For example, above the peak frequency of $\sim$3 GHz, the flux density has dropped by approximately 0.3 Jy over the course of the observations.  Conversely, below the peak frequency, the flux density has risen by approximately 0.5 Jy at 1.2 GHz.  These variations are at the $\sim$10\% level and are in opposite senses either side of the peak frequency.  The flux density at the peak frequency itself is less variable, rising by approximately 0.1 Jy over the course of the observations.  The spectral shape of PKS 1718$-$649 therefore evolves significantly with time.  These results support previous observations by \cite{gae00} that show that PKS 1718$-$649 is more variable than PKS 1934$-$638 over decade timescales.

Thus, the ATCA results presented in Figure 2 represent the best observations to date of the spectral variability (time variability of the shape of the radio spectrum) of PKS 1718$-$649 or indeed any member of the GPS/CSO classes of radio source.  With these measurements in hand, we can seek to interpret the spectral variability in terms of different models for absorption and emission processes in these objects.  This will be the primary subject of sections 3 and 4.

In addition to the ATCA results, we have further information on the PKS 1718$-$649 spectrum at lower frequencies, discussed in the remainder of section 2.  This additional information is also discussed in sections 3 and 4.

\subsection{Murchison Widefield Array observations and data analysis}
The MWA is a new low frequency interferometric radio telescope operating in the 80 - 300 MHz frequency range \citep{tin13,lon09}.  We have used the MWA to measure the flux density of PKS 1718$-$649 at a centre frequency of 199 MHz, using a 30 MHz bandwidth at each of the two linear polarisations.  A ``snapshot'' observation of 2 minutes duration was made on 2014 March 21 using all 128 antenna tiles of the array.  A calibration scan of Hercules A was also undertaken, immediately prior to the PKS 1718$-$649 observations.  

The data were analysed in \textsc{miriad}, as follows.  The auto-correlation spectra for the tiles were examined and three tiles were found to have anomolously high power levels, due to instrumental reasons.  The data for these tiles (numbered 56, 99, and 106 in the array) were flagged.  The task {\tt mfcal} was run on the remaining 125 tiles, on the Hercules A data, to derive bandpass corrections and an initial estimate of the tile-based gains, for each linear polaristaion independently.  These calibration parameters were copied to the PKS 1718$-$649 field data using {\tt gpcopy} and the data were imaged using {\tt invert} (4096$\times$4096 pixel images; 32.2\arcsec pixel sizes; robust=0.5 weighting; and multi-frequency treatment of the channelised visibilities).  The resulting images are dominated by a bright radio galaxy, far away from PKS 1718$-$649, PKS 1610$-$607, which is several hundred Jy at 200 MHz (see McKinley et al., in preparation, for a detailed study of PKS 1610$-$607 with the MWA), generating complex sidelobes across the field.  Initially only PKS 1610$-$607 was cleaned (task {\tt clean}), with subsequent self-calibration (task {\tt selfcal}) of phase and amplitude (only a single amplitude gain per antenna tile for the duration of the 2 minute observation) based on that model.  These actions substantially improved subsequent inversions of the incrementally calibrated visibilities, allowing images of good quality to be made at each linear polarisation, with PKS 1718$-$649 and many other known objects clearly detected.  After a full-field clean of 10000 iterations, the residual images were dominated by residual sidelobe confusion from PKS 1610$-$607 and imperfect modelling of the nearby Galactic Plane.  The images at each linear polarisation were independently corrected for the MWA primary beam response, which is different for each polarisation, and summed in the image plane to form a Stokes I image.  The RMS in the final image was approximately 170 mJy.

In order to examine the flux density scale in the vicinity of PKS 1718$-$649, a number of known radio sources within approximately 2$^{\circ}$ of PKS 1718$-$649 were inspected.  These objects are unresolved at MWA angular resolution.  The flux densities for these sources were measured from the MWA image by measuring the peak brightness and assuming a point source brightness distribution.  The measured flux densities were compared to cataloged data at 4.85 GHz from the Parkes-MIT-NRAO (PMN) survey \citep{gri93} and at 843 MHz from the Sydney University Molonglo Sky Survey (SUMSS) survey \citep{boc99,mau03}.  The spectral indices calculated from the 843 MHz and 4.85 GHz data, assuming a power law spectrum of the form $S\propto\nu^{\alpha}$, were extrapolated to 199 MHz to calculate an expected flux density, to be compared to the measurements from the MWA data.  Further, we take into account the errors on the published flux densities at 843 and 4850 MHz, to calculate errors on the spectral indices and hence errors on the predicted flux densities at 199 MHz.  The errors on the MWA measurements are taken as the combination in quadrature of the image RMS and an estimated error on the primary beam correction of approximately 10\% \citep{pbref}.

Table 3 contains the results of this analysis.  Of the 11 objects inspected, we find that 8/11 have measured 199 MHz flux densities that agree with the predicted 199 MHz flux densities to within the observational errors.  One object, PMN J1724$-$6443 has a measured flux density significantly higher than predicted, which may plausibly be the result of temporal variation of the flux density, given its relatively flat spectral index.  Two objects, PKS 1726$-$656 and PMN J1734$-$6421 have measured flux densities significantly lower than predicted, perhaps due to a flattening of the spectrum at low frequencies due to absorption processes.  Overall, 6/11 objects have flux densities higher than predicted (but generally within the errors) and 5/11 have flux densities lower than predicted (but again generally within the errors).  Thus, within the observational errors, there is no evidence to suggest a significant overall offset in the flux density scale in this image.  

The measured 199 MHz flux density for PKS 1718$-$649 is 1610 mJy.  Using the approach described above, the error on this measurement is approximately 240 mJy.

\begin{table}[ht]
\begin{center}
\begin{tabular}{ l c c c c} \hline
Source  & $S_{843}$             & $\alpha^{4850}_{843}$     & $S_{e,199}$          & $S_{m,199}$ \\
&(mJy)&&(mJy)&(mJy) \\ \hline
PMN J1705$-$6516 & $261.6\pm10.4$ & $-0.81\pm0.12$ & $840\pm_{160}^{200}$ & $940\pm190$ \\
PMN J1706$-$6453 & $212.3\pm6.5$ & $-1.08\pm0.21$ & $1010\pm_{290}^{400}$ & $1230\pm210$ \\
PKS 1708$-$648 & $568.8\pm17.1$ & $-1.09\pm0.09$ & $2740\pm_{410}^{470}$ & $3150\pm360$ \\
PMN J1715$-$6524 & $368.5\pm11.1$ & $-1.06\pm0.12$ & $1700\pm_{310}^{380}$ & $1540\pm230$ \\
PMN J1720$-$6358 & $263.3\pm7.2$ & $-0.94\pm0.14$ & $1020\pm_{210}^{260}$ & $630\pm180$ \\
PMN J1720$-$6452 & $196.8\pm7.7$ & $-0.86\pm0.16$ & $680\pm_{160}^{210}$ & $800\pm190$ \\
PMN J1724$-$6443 & $103\pm3.7$ & $-0.49\pm0.16$ & $210\pm_{50}^{60}$ & $740\pm190$ \\
PKS 1726$-$656 & $998.1\pm30$ & $-0.86\pm0.05$ & $3450\pm_{340}^{370}$ & $2610\pm310$ \\
PMN J1728$-$6432 & $186.1\pm8.8$ & $-1.04\pm0.23$ & $840\pm_{260}^{380}$ & $730\pm190$ \\
PMN J1734$-$6407 & $337\pm10.2$ & $-0.8\pm0.09$ & $1070\pm_{160}^{190}$ & $1290\pm210$ \\
PMN J1734$-$6421 & $950.9\pm28.6$ & $-0.95\pm0.06$ & $3750\pm_{410}^{460}$ & $1670\pm240$ \\
\end{tabular}
\caption{Data for 11 radio sources in the vicinity of PKS 1718$-$649.  $S_{843}$ is the flux density at 843 MHz from the SUMSS catalog, in mJy.  $\alpha^{4850}_{843}$ is the spectral index between 843 MHz and 4850 MHz from the SUMSS and PMN catalogs.  $S_{e,199}$ is the expected flux density at 199 MHz, extrapolated from 843 and 4850 MHz, in mJy.  $S_{m,199}$ is the 199 MHz flux density measured from the MWA image, in mJy.}
\end{center}
\end{table}

While this flux density estimate is approximate, for the reasons given above, it nevertheless suits our purposes well.  As can be seen in section 3, different absorption models give very different predictions for the flux density in the MWA frequency range, with one family of models predicting mJy flux densities and another family of models predicting Jy flux densities.  Thus, even an approximate estimate of the flux density from the MWA data is very useful, as it clearly supports one family of models over another on a qualitative basis.  Since the MWA data are not simultaneous with the ATCA data, we do not use the MWA data explicitly in the spectral modelling for PKS 1718$-$649.

\subsection{Parkes observations and data analysis}
Observations of PKS 1718$-$649 were made with the Parkes radio telescope on 2014 April 9, at a centre frequency of 732 MHz and with a bandwidth of 64 MHz.  During the observation, local Digital TV transmitters were active, with the result that much of the band was lost to interference.  After flagging affected data, a 6.5 MHz usable bandwidth remained, centred on 725.75 MHz.  The flux density of PKS 1718$-$649 was measured in this band, calibrated by observations of PKS 1934$-$638 (same calibrator as for the ATCA observations described above), yielding a flux density of 3.84$\pm$0.19 Jy.  With the large beam of the Parkes radio telescope at this frequency ($\sim$30 arcminutes), it is possible that this measurement is affected by diffuse galactic synchrotron variations in this region of the sky.  However, we expect any error due to this effect to be small compared with the approximate 5\% error on the flux density scale that dominates the quoted error on the measurement.  The Parkes data are included in the plotted spectra in Figure 2.

As above for the MWA data, the Parkes data have not been used explicitly in the spectral modelling, as the Parkes data are not simultaneous with the ATCA data.  However, along with the MWA data, the Parkes data are useful in qualitatively assessing the plausibility of different families of absorption models.

\section{Discussion}
\subsection{Application of absorption models to the PKS 1718$-$649 spectrum}
\cite{tin03} explored a wide range of models involving free-free absorption and synchrotron self-absorption as explanations for the peaked radio spectrum of PKS 1718$-$649.  The most successful model in terms of its fit to the data was that adopted from \cite{bic97}.  We have adopted a similar approach here, attempting fits of the same models to these improved data.  

The absorption model fits to the spectra of PKS 1718-649 were performed using a non-linear least squares routine that applied the Levenberg-Marquardt algorithm, an iterative procedure that linearises the function at each step based on a new estimate of the function from the gradient of the previous step. The minimum of the function is found when the derivative of the sum of squares with respect to new parameter estimates is zero. To ensure that the fitting routine finds the global minimum, and not a local minimum, different initial parameters were used to ensure the fitting routine converged to the same parameter values. 

The covariance matrix produced by the fitting routine was formed by multiplying the Jacobian approximation to the Hessian of the least squares objective function by the residual variance. The reported uncertainties on the parameters to the fits were taken from the diagonal terms of this covariance matrix and represent the 68.27\% confidence interval (one-sigma). The uncertainties on the data points were assumed to be Gaussian for this analysis.

The fitting has only been undertaken using the ATCA data, as neither the MWA nor Parkes data were obtained simultaneous with the ATCA observations.  The MWA and Parkes data are plotted along with the ATCA data, but have not been used in the fits.

Figure 2 shows the results of the various fits.  As can be seen, Figure 2 reproduces many aspects of the \cite{tin03} results in that all of the homogeneous free-free and synchrotron self-absorption models drop too quickly at low frequencies, with a large discrepancy at the lowest frequencies covered by Parkes and the MWA.  Additionally, Figure 2 reproduces another result of \cite{tin03}, that the most successful model is the inhomogeneous free-free absorption model of \cite{bic97}.  The results of the application of the \cite{bic97} model to the ATCA data are listed in Table 4.  The Parkes and MWA data broadly agree with the form of the low frequency power law spectrum predicted by this model.  

\begin{table}[ht]
\begin{center}
\begin{tabular}{ l c  c c c c} \hline
Date  & $S$ (Jy) & $\alpha$      & $p$                   & $\nu_{0}$ (GHz) & Reduced $\chi^2$   \\ \hline
Feb 12&7.274$\pm$0.002&0.716$\pm$0.005&$-$0.421$\pm$0.004&3.840$\pm$0.004&0.45 \\
Feb 13&7.077$\pm$0.002&0.747$\pm$0.004&$-$0.457$\pm$0.004&3.855$\pm$0.003&1.04 \\
Dec 13&6.758$\pm$0.007&0.746$\pm$0.001&$-$0.498$\pm$0.001&4.052$\pm$0.010&0.72 \\ \hline
\end{tabular}
\caption{The parameters of the \cite{bic97} model as fitted to the PKS 1718$-$649 data.}
\end{center}
\end{table}

The \cite{bic97} model continues to provide an excellent description of the PKS 1718$-$649 radio spectrum, even with the addition of measurements below the ATCA frequency range and a greatly improved frequency coverage (particularly below the peak frequency).  In the next section we examine the spectral variability of PKS 1718$-$649 in terms of both free-free and synchrotron self-absorption explanations.

One of the main constraints from the data on the absorption mechanisms under consideration in the next section is the value of the peak frequency.  From a Markov Chain Monte Carlo analysis of the \cite{bic97} model fits to the data, using the model parameter errors in Table 4, the peak frequency and one sigma error on the peak frequency at each epoch were estimated: $\nu_{\rm p}$=3.399$\pm$0.001 GHz for February 2012; $\nu_{\rm p}$=3.175$\pm$0.001 GHz for February 2013; and $\nu_{\rm p}$=3.163$\pm^{0.001}_{0.002}$ for December 2013.  These results are model-dependant and slightly different results may be obtained if models other than the \cite{bic97} model are used.  However, given the quality of the fits indicated by the reduced $\chi^2$ values listed in Table 4, the model appears to adequately characterise the data such that the estimates of peak frequency can be trusted.

\subsection{Interpretation of spectral variability}
The spectral variability observed in PKS 1718$-$649 is most likely intrinsic.  Several arguments suggest that the contribution of interstellar scintillation is negligible.  Specifically, interstellar scintillation at centimetre wavelengths is only present if the source exhibits structure on scales of less than tens of microarceseconds.  However, 22 GHz VLBI imaging \citep{tin03} reveals that at most 10 mJy of the emission is unresolved, and most of the emission occurs on scales of ~5 mas, which is too large to be subject to the effects of interstellar scintillation.  The structure is likely even more extended at frequencies $<10\,$GHz, since it has a steep spectrum.  Moreover, inspection of Figure 2 shows that the variations between epochs does not resemble the frequency dependence characteristic of interstellar scintillation \citep{Narayan}. For a source observed off the Galactic plane, the amplitude of scintillations peaks in the range $\sim 3-6\,$GHz and is typically less than 50\% of this value at 1.4\,GHz and 10\,GHz \citep{Wa98,Kedziora-Chudczer}.  Given this conclusion, the remainder of this discussion examines the spectral variability of PKS 1718$-$649 in terms of mechanisms intrinsic to the AGN.

Based on the relatively compact nature of PKS 1718$-$649 revealed by the VLBI data presented in \cite{tin03} and given that they were unable to ascertain the precise origin of the spectral variability, \cite{tin03} concluded that synchrotron self-absorption was the most likely explanation for the peaked spectrum of PKS 1718$-$649, given the turnover frequency.  However, with better data available, and solid evidence for spectral variability in PKS 1718$-$649, we can revisit the question of the cause of the peaked spectrum in this object.

If the turnover were due to synchrotron emission, then the changes in the turnover frequency (i.e. between approximately $\nu_{\rm p} = 3.163$\,GHz and $\nu_{\rm p}=3.399\,$GHz) would be accompanied by changes in the peak flux density that are considerably larger than the $0.12\,$Jy observed.  The peak flux density, $S_{\nu,t}$ at the peak of the turnover, $\nu_{\rm p}$, in a simple synchrotron model is derived by equating $k_B T_B$, where $T_B$ is the rest-frame brightness temperature of the emission, to the energy of the emitting particles:
\begin{eqnarray}
k_B T_B \approx m_e c^2 \left( \frac{2 \pi m_e \nu_{\rm p}}{0.47 e B \sin \theta}  \right)^{1/2},
\end{eqnarray}
where $\theta$ is the angle that the magnetic field of strength $B$ makes to the line of sight and $k_B$, $e$ and $m_e$ are the Boltzmann constant, electron charge, and electron mass respectively.  For a source of angular diameter $\theta_S$, one therefore has,
\begin{eqnarray}
S_{\nu,t} = \left( \frac{\pi^3 m_e^3 \nu_{\rm p}^5 \theta_S^4}{0.94 e B \sin \theta} \right)^{1/2},
\end{eqnarray}
and the measured variation in $\nu_{\rm p}$ would imply an approximate 20\% (i.e. $\sim$1.1\,Jy) variation in the peak flux density in the context of this model.  Since such variations of this order are not observed (only $\sim$10\% of this level is observed), it is difficult to straightforwardly attribute the observed changes in the spectral turnover frequency to changes in the synchrotron optical depth.  We note that a change in $\nu_{\rm p}$ could instead be caused by changes in $B \sin \theta$ or $\theta_S$, however  $\nu_{\rm p}$ is relatively insensitive to changes in $B$: the observed change in $\nu_p$ would require a 30\% increase in $B \sin \theta$ for fixed $S_{\nu,t}$.  If the change in $\nu_{p}$ were instead ascribed to source size changes, it would require a 7\% {\it contraction} of the source diameter, which appears implausible.

In the context of the free-free absorption models, other comments on the spectrum deserve mention.  The $\sim 0.5\,$Jy flux density variations observed below the spectral turnover (e.g. at 1\,GHz) could be due to variations in the free-free opacity, caused by the passage of individual clouds (in an inhomogeneous medium) across the line-of-sight to a compact component of the source.  The free-free opacity scales as $\nu^{-2.1}$ \citep{alt60,lang80}, so such inhomogeneity could explain at most $0.01\,$Jy of the $\sim 0.3\,$Jy flux density changes observed at $\nu = 4.5\,$GHz and only 3\,mJy of the similar variation at 9\,GHz.  It therefore seems that a model beyond the context of free-free models is required to explain the variability above the peak frequency.

We can place limits on the free-free opacity variation.  The flux density difference, $\Delta S$, for an opacity that changes from $\tau_{\rm ff}$ to $\tau_{\rm ff} + \Delta \tau_{\rm ff}$ is $\Delta S= S_0 e^{-\tau_{\rm ff}} (1 - e^{- \Delta \tau_{\rm ff}})$, where $S_0$ is the flux density of the compact component covered by the absorbing material.  Hence,
\begin{eqnarray}
\Delta \tau = - \ln \left[1 - \frac{\Delta S}{S_0 e^{-\tau_{\rm ff}}} \right].
\end{eqnarray}
Assuming that $\tau_{\rm ff} > 0$ (i.e. absorption only), the observed 0.5\,Jy flux density variation at 1\,GHz constrains the opacity change to be $\Delta \tau > 0.13$ for $S_0=4.0\,$Jy.  

This opacity change would imply the existence of large density gradients in the absorbing medium.  Taking a fiducial speed for the absorbing clouds of $v=200 v_{200}$\,km\,s$^{-1}$, a variation on a timescale of $T$ years traces structure on a scale of $2 \times 10^{-4}\,T \,v_{200}\,$pc, and the implied gradient is $\Delta \tau_{\rm ff}/\Delta L = 4.9 \times 10^3 \Delta \tau\, T^{-1} v_{200}^{-1}\,$pc$^{-1}$.  Using the approximation for the free-free opacity used in \cite{bic97} and derived in \cite{alt60}, this in turn implies an emission measure gradient of
\begin{eqnarray}
\left( \frac{\Delta {\rm EM}}{\Delta L} \right) = 1.5 \times 10^{10} \Delta \tau\, T^{-1} v_{200}^{-1} \left( \frac{T_{\rm gas}}{10^4\,{\rm K}} \right)^{1.35}  \left( \frac{\nu}{1\,{\rm GHz}}\right)^{2.1}  {\rm cm}^{-6}{\rm \,pc\,pc}^{-1}. 
\end{eqnarray}

The idea of a rapidly variable column of ionised material providing a highly variable free-free opacity is qualitatively supported by the observations of ubiquitous variability in the X-ray absorbing columns of neutral hydrogen in Seyfert 2 galaxies by \cite{ris02}.  These authors studied 25 X-ray defined Seyfert 2 galaxies and found that variations in the neutral hydrogen X-ray absorbing column were present in 22/25 of the galaxies at the 20\% to 80\% level.  The timescales for variability were less than one year.  \cite{ris02} suggest that these results rule out a homogenous obscuring torus and support the presence of a clumpy circumnuclear medium with structure scales well below 1 pc.  If variations of this nature are seen in neutral hydrogen, we suppose it is reasonable that similar variations will be present in the ionised component of the circumnuclear environment, especially if the environment is fragmented into an ensemble of sub-pc scale clouds.

For example, a uniform electron/ion density of $3\times10^{4}$ cm$^{-3}$ over a path length of 10 pc gives an EM of $\sim10^{10}$ cm$^{-6}$pc.  A 10\% variation in this EM over a length scale of $\Delta L$ = 0.1 pc gives an EM gradient of the same order of magnitude as equation 4.  For these parameters, assuming only hydrogen atoms and an ionisation fraction of 10\%, the implied neutral hydrogen column density is $10^{25}$ cm$^{-2}$.  This column density is within an order of magnitude of the highest column densities considered by \cite{ris02}.  Given the level of uncertainty in most of the parameters in equation 4, and the assumptions above, this represents reasonable consistency.  In addition, the results of \cite{ris02} show that the timescales of variability in the column density are in the range of months to years, consistent with the variability timescales seen below the peak frequency in the radio.  The results of monitoring PKS 1718$-$649 at 843 GHz by \cite{gae00} show time variations consistent with the ATCA data in the current study and the proposed time-scale of neutral hydrogen column density variations.

Within the context of the \cite{bic97} model for free-free absorption, the discussion above would relate to changes in the number and size/density distributions of absorbing clouds along the line of sight to the radio source.  A requirement of the \cite{bic97} model is that whatever variations in the absorbing clouds take place, the distribution must remain in the form of a power law.

In principle, some of the evolution observed in the PKS 1718$-$649 spectra could be attributed to various cooling mechanisms.  The effect of three forms of cooling on the spectral evolution of the emission from the hotspot of a compact source was considered by \cite{man02}.  The principal mechanisms are: adiabatic expansion and cooling; synchrotron cooling; and Inverse Compton cooling.  While these three mechanisms do not adequately explain the variability across the entirety of the PKS 1718$-$649 spectrum, they may explain some facets of its variability, particularly in the optically thin region of the spectrum.  However, the synchrotron and Inverse Compton cooling models fail in detail because there is no evidence in the data for a break in the optically thin part of the spectrum.  Both synchrotron and Inverse Compton cooling insert a break in the spectrum that changes the slope (e.g. from $\alpha \rightarrow \alpha+1/2$) but there is no evidence for a break in the optically thin part of the spectrum observed for PKS 1718$-$649.  A break may exist at frequencies higher than our 10 GHz upper frequency.

It is possible that adiabatic expansion plays some role in the time evolution of the spectrum, however it cannot explain the observed changes in spectral shape with time.  For a source expanding isotropically, the electron energy scales as the linear dimension {\bf ($L$)} of the source. For a particle with energy $\epsilon$, we have $\epsilon \propto L^{-1}$ and for a source expanding at a constant speed $\epsilon = \epsilon_0 (t_0/t)$, the power law distribution of electron energies at some time $t$ after $t_0$ is {\bf (Shklovskii 1960)}:
\begin{eqnarray}
N(\epsilon,t) = K_0 \left( \frac{t}{t_0} \right)^{-(2 \alpha + 3)} \epsilon^{-(2 \alpha+1)}.
\end{eqnarray}  

This then implies that the spectrum of the source retains its initial shape but its flux density scales with time as $S_\nu \propto t^{-2 (2 \alpha+1)}$.  The upper and lower cutoffs in the energy distribution similarly decrease.  In summary, it is therefore possible in principle to attribute changes in the amplitude of the optically thin emission of the spectrum of PKS 1718$-$649 to adiabatic losses, but it is not possible to explain the evolution of the entire spectrum using this mechanism.

A combination of adiabatic losses in the optically thin part of the PKS 1718$-$649 spectrum and variability of the Emission Measure causing the spectral turnover and variability of the low frequency spectrum, appears to be a plausible explanation for the shape and variation of the PKS 1718$-$649 spectrum.

A possible route to model both the synchrotron spectrum above the spectral peak and the apparently free-free absorbed synchrotron spectrum below the spectral peak may be via numerical simulations of hotspots, where jets impinge upon an ensemble of clouds, such as in the dense and kinematically confused AGN environment \citep{bic07}.  \cite{sax02}  present a simulation of time-variability in AGN hotspots.  The variability evident in their simulations appears to be due to the surging of the jet near the hot-spot caused by alternate compression and decompression of the jet by turbulence in the backflow in the cocoon.  The dynamical timescale depends on the ratio of the jet diameter to the internal sound speed.

However, the simulations are of limited applicability to the present problem because there is no straightforward means of deducing the synchrotron spectrum on the basis of the underlying density field. The model is purely hydrodynamical, and the absence of magnetic field information impedes detailed investigation of the emissivity of the jet.  In the simulation the emissivity is deduced by equating the magnetic pressure to the particle pressure, assuming a specific magnetic field geometry and assuming that the emission is optically thick.   The absence of opacity effects in their simulation precludes its application to PKS 1718$-$649.

\section{Summary}
We have presented new measurements of the radio spectrum of PKS 1718$-$649 between the frequencies of 1 and 10 GHz, using the Compact Array Broadband Backend of the Australia Telescope Compact Array at three epochs spread over approximately 21 months.  These measurements improve on previous attempts to measure the radio spectrum of this object by \cite{tin03}.  The new data show that significant spectral variability is present for this object, which is interesting when considering the absorption mechanisms that could produce the spectral turnover observed.  The free-free absorption model of \cite{bic97} provides the best fit of the spectral data at each epoch of observation, as was the case for \cite{tin03}, and remains the best model when flux density measurements below the ATCA frequency range are considered, obtained at 726 MHz from Parkes and at 199 MHz from the Murchison Widefield Array.  When considering the spectral variability of PKS 1718$-$649, neither synchrotron self-absorption nor free-free absorption can explain all aspects of the variability.  A model in which adiabatic losses in a synchrotron spectrum account for the variability in the optically thin part of the spectrum above the peak frequency, and variations in the free-free opacity cause variations at frequencies below the peak, appears to be plausible.  This is supported by observations of large variations in the neutral hydrogen column density toward a number of AGN from X-ray observations.  As a typical example of a Compact Symmetric Object, this analysis may hold important conclusions for the CSO class as a whole.  Similar monitoring observations of spectral variability in a large sample of similar objects would therefore be interesting as future work.

\acknowledgements
This scientific work makes use of the Murchison Radio-astronomy Observatory, operated by the Commonwealth Scientific and Industrial Research Organisation (CSIRO). We acknowledge the Wajarri Yamatji people as the traditional owners of the Observatory site. Support for the MWA comes from the U.S. National Science Foundation (grants AST-0457585, PHY-0835713, CAREER-0847753, and AST-0908884), the Australian Research Council (LIEF grants LE0775621 and LE0882938), the U.S. Air Force Office of Scientific Research (grant FA9550-0510247), and the Centre for All-sky Astrophysics (an Australian Research Council Centre of Excellence funded by grant CE110001020). Support is also provided by the Smithsonian Astrophysical Observatory, the MIT School of Science, the Raman Research Institute, the Australian National University, and the Victoria University of Wellington (via grant MED-E1799 from the New Zealand Ministry of Economic Development and an IBM Shared University Research Grant). The Australian Federal government provides additional support via the CSIRO, National Collaborative Research Infrastructure Strategy, Education Investment Fund, the Australia India Strategic Research Fund, and Astronomy Australia Limited, under contract to Curtin University. We acknowledge the iVEC Petabyte Data Store, the Initiative in Innovative Computing and the CUDA Center for Excellence sponsored by NVIDIA at Harvard University, and the International Centre for Radio Astronomy Research (ICRAR), a Joint Venture of Curtin University and The University of Western Australia, funded by the Western Australian State government.  SJT acknowledges support from the Western Australian Government via a Premier's Research Fellowship.  The Australia Telescope Compact Array is part of the Australia Telescope National Facility which is funded by the Commonwealth of Australia for operation as a National Facility managed by CSIRO.  The Parkes Radio Telescope is part of the Australia Telescope National Facility which is funded by the Commonwealth of Australia for operation as a National Facility managed by CSIRO.
{\it Facility:} \facility{ATCA, MWA, Parkes}.

\end{document}